\include{psfig}
\documentstyle[11pt]{article}
\newcommand{\beq}{\begin{equation}}
\newcommand{\eeq}{\end{equation}}
\newcommand{\beqa}{\begin{eqnarray}}
\newcommand{\eeqa}{\end{eqnarray}}
\renewcommand{\d}{{\mathrm d}}

\renewcommand{\b}{\tilde{b}}
\renewcommand{\le}{\stackrel{{\textstyle <}}{_ {\displaystyle \sim}}}

\newcommand{\pbar}{\overline{p}}
\newcommand{\qbar}{\overline{q}}
\newcommand{\cbar}{\overline{c}}
\newcommand{\Cbar}{\overline{C}}
\newcommand{\A}{{\mathcal A}}
\renewcommand{\P}{{\mathcal P}}
\newcommand{\T}{{\mathcal T}}
\newcommand{\R}{{\mathcal R}}
\begin{document}

\baselineskip 18pt

\newcommand{\sheptitle}
 { Wide-angle elastic scattering and color randomization}

\newcommand{\shepauthor}
 {Michael G. Sotiropoulos }

\newcommand{\shepaddress}
 {Physics Department, University of Southampton\\
  Southampton, SO17 1BJ, U.K.}

\begin{titlepage}

\begin{flushright}
SHEP 95/39\\
hep-ph/9512397\\
December 1995
\end{flushright}

\vspace{.4in}

\begin{center}
{\large{\bf \sheptitle}}
\bigskip \\
\shepauthor \\
\mbox{} \\
{\it \shepaddress} \\
\end{center}

\vspace{.5in}

\begin{abstract}

 Baryon-baryon elastic scattering is considered in the
 independent scattering (Landshoff) mechanism.
 It is suggested that for scattering at moderate energies,
 direct and interchange quark channels contribute with equal
 color coefficients because  the quark color  is randomized
 by soft gluon exchange during the hadronization stage.
 With this assumption, it is shown that the ratio of cross sections
 $R_{\overline{p} p/ p p}$  at CM angle $\theta = 90^0$
 decreases from a high energy value of
 $R_{\pbar p / pp} \approx 1/2.7$, down to
 $R_{\pbar p / pp} \approx 1/28$,
 compatible with experimental data at moderate energies.
 This sizable fall in the ratio seems to be characteristic
 of the Landshoff mechanism, in which changes at the quark level
 have a strong effect precisely because the hadronic process occurs
 via multiple quark scatterings.
 The effect of color randomization on the angular distribution of
 proton-proton elastic scattering and  the cross section ratio
 $R_{np/pp}$ is also discussed. \\

 \noindent
 PACS numbers: 13.60.Fz, 13.85.Dz  \hfill

\end{abstract}

\setcounter{page}{0}

\begin{flushleft}
{\footnotesize
--------------------------------------------------- \\
E-mail: mgs@hep.phys.soton.ac.uk  }

\end{flushleft}

\end{titlepage}

\newpage

\section{Introduction}

 The analysis of exclusive hadronic processes within the framework
 of perturbative QCD (pQCD) remains conceptually as well as
 computationally challenging. In the case of hadronic elastic scattering
 at high energy and wide CM angle, the formalism
 for the treatment of the amplitude has been developed in the
 kinematic region of momentum transfer $|t|$ much larger than
 the hadronic mass scales. The long distance dynamics of the
 hadronic bound state factorize from the short distance scattering
 of the constituent quarks.
 Specifically, for baryon-baryon elastic scattering the amplitude
 takes the factorized form \cite{BL}
 \beq
 \A (s,t; h_i) =
 \int \prod_{i=1}^4 [ \d x_i ] \phi_i([x_i],[\lambda_i],h_i;\mu)
 M_H( \hat{s}_{i j}, [\lambda_i]; \mu)  \, .
 \label{factor}
 \eeq
 The hard scattering amplitude $M_H( \hat{s}_{i j}, [\lambda_i]; \mu)$
 describes the scattering of nearly collinear constituent quarks
 with helicities $[\lambda_i]$.
 It depends on the quark invariants $\hat{s}_{i j}$ but not on the
 hadronic mass scales.
 At lowest order in $\alpha_s$, $M_H$ is equal to the Born amplitude
 with, in principle, calculable higher order corrections.
 The quark distribution amplitudes $\phi_i([x_i], [\lambda_i],h_i; \mu)$
 describe the three valence quark component of the baryon wavefunction
 with helicity $h_i$ and it is
 evaluated at factorization scale $\mu^2 = {\mathcal O}(|t|)$,
 with calculable $\ln t$ corrections.
 In the single hard scattering mechanism, where all constituents
 scatter together in a small space-time region,
 the calculation of $M_H$ at lowest order,
 ${\mathcal O}(\alpha_s^5(\mu))$,  requires the
 evaluation of approximately 300,000 distinct tree graphs for
 $6 q \rightarrow 6 q$ scattering \cite{FarrarNeri}.
 Moreover, the inclusion of higher order radiative corrections
 and the implementation of the factorization of infrared singularities
 make this approach unyielding.
 Alternatively, two distinct scattering mechanisms have been
 considered.
 The first is the quark interchange model (QIM) \cite{Gunion},
 in which the scattering is assumed to proceed via the exchange of a
 pair of quarks between the scattering hadrons.
 The second is the independent scattering (Landshoff) mechanism
 \cite{Land}, in which the quarks from each initial hadron
 scatter pairwise and independently up to logarithmic radiative
 corrections.
 From the pQCD point of view both QIM and Landshoff-type
 diagrams originate as particular pinch singularities
 of the single hard scattering diagrams, although the two sets are
 distinct \cite{RamseySivers1}.

 The experimental studies of wide-angle elastic scattering at moderate
 energies suggest that the process is mainly driven by the QIM
 mechanism \cite{Blazey, AGS}.
 In particular, QIM is consistent with the dimensional
 counting scaling behavior of the elastic baryon cross section
 \beq
 \frac{\d \sigma^{BB}}{\d t} \sim \frac{1}{s^{10}} f(\theta) \, ,
 \label{scaling}
 \eeq
 its dependence on the CM angle $\theta$
 as given by the function $f(\theta)$,
 as well as its flavor and crossing properties manifested by the
 cross section ratios such as
 \beq
 R_{\pbar p / p p}(s,\theta) =
 \frac{\d \sigma^{\pbar p}/ \d t}{\d \sigma^{pp}/ \d t} \, .
 \label{Rdef}
 \eeq
 On the other hand, the Landshoff mechanism seems to have negligible
 contribution to the elastic scattering.
 This fact remains a puzzle since, within pQCD,
 independent quark scattering,
 being ${\mathcal O}(\alpha_s^3(\mu))$ modulo radiative corrections,
 is anticipated to contribute.
 Indeed, Botts \cite{Botts} has studied numerically
 the Landshoff mechanism with Sudakov resummed radiative corrections.
 With reasonable choices for the end-point and
 infrared cutoff parameters he has concluded that
 Landshoff-type contributions to the cross section are non-negligible
 and must be included in the phenomenology of elastic scattering.

 In this paper we consider the above puzzle.
 Specifically we reexamine the Landshoff mechanism and study
 to what extent it can account for
 certain features of the elastic baryon cross sections such as
 their angular dependence and the ratios
 $ R_{\pbar p/pp}(s,\theta)$ and
 $ R_{n p/p p}(s,\theta) $
 in the moderate energy regime where measurements are available.
 Recent experiments at AGS \cite{AGS} have measured
 $R_{\pbar p/p p} \approx 1/40$ at $\sqrt{s}=3.59$~GeV and  $\theta=90^0$.
 This measurement is near the beginning of the scaling region
 as given by eq.~(\ref{scaling}).
 Of course the energy here is not high enough for
 a fully self-consistent perturbative treatment
 of the process in terms of independent hard scatterings.
 Nevertheless, in a constituent quark model, $R_{\pbar p/ p p}$ is largely
 determined by the flavor flows, i.e. the number of possible routings
 of the quarks among the participating hadrons, and the color
 factors arising from combining the color structure of the hard scatterings
 with the color singlet external hadrons.
 Expecting this ratio to be less sensitive than
 the elastic cross section itself to the factorization assumptions,
 we compute it using the formalism of pQCD.
 In a sense the treatment presented below supplies a QCD-motivated model which
 realizes the observation that elastic scattering is dominated
 by quark interchange.

 Our starting point is the factorized form of the elastic amplitude
 in the Landshoff mechanism, which we briefly review in section 2.
 In section 3 the idea of color randomization is presented
 and its effect on the crossing properties of the elastic amplitude
 is illustrated by considering a toy model of scalar quarks.
 The main point is that there is  always
 soft gluon exchange among the constituent
 quarks in the initial and final state
 which cannot be factored into the hadronic wave functions and
 mixes the quark color.
 In the asymptotic high energy regime, where soft radiation can
 be treated perturbatively, it is possible to relate the color
 of the quarks at the hadronization region with their color
 at the hard scattering region by computing color traces order
 by order in $\alpha_s$.
 But for moderate energies, where a perturbative expansion
 for soft gluon exchange is not self-consistent, we suggest that the
 effect of soft radiation is to decorrelate the color of the
 constituent  quarks at hadronization from the color they have when
 they participate in the hard scatterings.
 In other words, by the time the quarks enter the hard scattering
 region their color has been randomized by soft gluon exchange.
 We express this color randomization by requiring that all
 quark channel combinations, i.e. direct ($ttt$), total interchange ($uuu$),
 single and  double interchange ($ttu$, $tuu$),
 contribute with the same color coefficients
 when summed in the calculation  of the elastic proton-proton amplitude.
 In section 4 we compute the $pp$, $\pbar p$ and $n p$ elastic amplitudes
 in the helicity basis and at lowest order
 in $\alpha_s$ for the hard scatterings.
 At asymptotically high energies we expect color flow to be dominated by
 lowest order contributions.
 This gives an asymptotic ratio
 $R^{\mathrm as}_{\pbar p/p p} \approx 1/2.7$  at $\theta=90^0$.
 In the color randomization model suggested for subasymptotic energies
 we find  $R^{\mathrm rand}_{\pbar p/ p p} \approx 1/28$.
 The effect of color randomization on the angular distribution
 of proton-proton elastic scattering and the ratio
 $R_{np /pp}$ is also considered.
 We end by discussing these results.

\section{The Landshoff mechanism in pQCD}

 The structure of the elastic amplitude via independent quark scatterings
 \cite{Land}
 is  shown in fig.~1 for $pp \rightarrow pp$.
 Only the three valence quark part of the proton wave function is considered
 and $M^1, M^2$ and $M^3$ represent  on-shell quark-quark scatterings.
 In leading twist factorization the hard scatterings $M^m$, $m=1,2,3,$
 depend only on the longitudinal quark momenta $x_{m,i} P_i$,
 that  scale with $\sqrt{s}$ in the proton CM frame.
 The longitudinal momentum fractions are characterized by both
 a hadronic label $i$ and a scattering label $m$.
 The kinematics of on-shell q-q scattering requires
 that quarks participating in the same hard scattering have equal
 momentum fractions, i.e.
 \beq
 x_{1,i} = x_1 \geq 0 \, ,
 \hspace{1cm}
 x_{2,i} = x_2 \geq 0 \, ,
 \hspace{1cm}
 x_{3,i} = 1-x_1 -x_2 \geq 0\, ,
 \label{xis}
 \eeq
 for every $i=1,2,3,4$  up to $O(1/ \sqrt{s})$ corrections.
 Dependence on transverse  momentum and hadronic mass scales resides
 in the hadronic wave functions.

 The hard scatterings lie along the spacelike direction $\eta^\mu$,
 perpendicular to  the scattering plane.
 This is the line of intersection of the Lorentz contracted
 wave functions of the incoming and outgoing protons.
 We denote by $b_m$ the positions of the
 hard scatterings $M^m$ along the $\eta^\mu$ direction
 and by $\b_m$ their  mutual transverse separations defined as
 \beq
 \b_1=b_2-b_3 \, ,
 \hspace{1cm}
 \b_2=b_1-b_3 \, ,
 \hspace{1cm}
 \b_3=\b_2-\b_1 \, .
 \label{btilde}
 \eeq

 The three-quark component of the proton wave function
 is obtained as a Fourier transform of the three-quark operator
 \cite{CZrep}
 \beqa
 Y_{\alpha \beta \gamma} (k_1, k_2; P ,h) &=&
 \frac{(\sqrt{2} E)^{1/2}}{ N_c!}
 \int \frac{ \d^4 y_1}{(2 \pi)^4} e^{ i k_1 \cdot y_1 }
      \frac{ \d^4 y_2}{(2 \pi)^4} e^{ i k_2 \cdot y_2 }
 \nonumber \\
 & & \times
 \langle 0 |
 T \left( u_\alpha^a(y_1) u_\beta^b(y_2) d_\gamma^c(0) \right)
 |P, h \rangle
 \epsilon_{a b c} \, ,
 \label{Y}
 \eeqa
 where $E$ is the energy of the fast moving proton and $h$ its
 helicity.
 The wave function is decomposed in terms of valence quark
 spinors with definite helicity.
 Defining the dimensionless structures \cite{SS2}
 \beqa
 &\ &{\cal M}^{(1)}_{\alpha \beta \gamma} =
 (E_1 E_2 E_3 )^{-1/2} \;
 u_\alpha(k_1,+) \; u_\beta(k_2,-)  \; d_\gamma(P-k_1-k_2,+) \, ,
 \nonumber \\
 &\ &{\cal M}^{(2)}_{\alpha \beta \gamma} =
 (E_1 E_2 E_3 )^{-1/2} \;
 u_\alpha(k_1,-) \; u_\beta(k_2,+) \; d_\gamma(P-k_1-k_2,+) \, ,
 \nonumber \\
 &\ &{\cal M}^{(3)}_{\alpha \beta \gamma} =
 - (E_1 E_2 E_3 )^{-1/2} \;
 u_\alpha(k_1,+) \; u_\beta(k_2,+) \; d_\gamma(P-k_1-k_2,-)  \, ,
 \label{helspinors}
 \eeqa
 where $E_1 ,\, E_2$ and $E_3$ are the energies of the two $u$-quarks
 and the $d$-quark respectively,
 we obtain the helicity decomposition of the wave function.
 In impact parameter ($\b$-) space and for $h=+$ this is
 \beq
 \tilde{Y}_{\alpha\beta\gamma}(x_1,x_2,x_3, \b_1,\b_2; h=+) =
 \frac{2^{1/4}} {8 N_c!}
 \bigg [
   \P_{123} {\cal M}^{(1)}_{\alpha \beta \gamma}
 + \P_{213} {\cal M}^{(2)}_{\alpha \beta \gamma}
 + 2 \T_{123}
 {\cal M}^{(3)}_{\alpha \beta \gamma}
 \bigg ] \, ,
 \label{tildeYexpand}
 \eeq
 where
 \beq
 \T_{123} \equiv \frac{1}{2} (\P_{132}+\P_{231}) \, ,
 \label{T}
 \eeq
 and $\P_{123} \equiv \P (x_1,x_2,x_3;\b_1,\b_2)$ is
 the proton wave function projected along the $\eta^\mu$-direction.
 Its dependence on the transverse separations $\b_m$ can be computed
 perturbatively
 via soft gluon resummation and results in a Sudakov exponent
 to be specified below.
 The connection of $\P$ to the usual light-cone
 distribution amplitude $\phi$ is given in perturbation theory via
 \beq
 \P(x_1, x_2, x_3,
       \b_1 \rightarrow 0, \b_2 \rightarrow 0; \mu) =
 f_N(\mu) \,  \phi(x_1, x_2, x_3; \mu) + {\mathcal O}(\alpha_s(\mu))
 \, ,
 \label{conn}
 \eeq
 where $f_N(\mu)$ is an overall normalization parameter,
 \beq
 f_N(\mu=1 \textrm{GeV}) = (5.2 \pm 0.3) \times 10^{-3} \textrm{GeV}^2
 \, .
 \label{fN}
 \eeq
 In the asymptotic energy limit $\left( \P_{\mathrm as} \right)_{123}$
 becomes symmetric upon permutation of its arguments.
 The asymptotic light-cone distribution amplitude is
\beq
\phi_{\mathrm as}(x_1, x_2, x_3) = 120 x_1 x_2 x_3 \, .
\label{phiasy}
\eeq
 For subasymptotic energies model dependent $\phi$'s  \cite{LCDA}
 are more suitable for reproducing the overall normalization
 of the exclusive process in which the proton participates.
 Finally, the color structure of the hadronic wave function is of the
 form $\epsilon_{a b c}$.

 The main feature of the Landshoff mechanism for elastic scattering
 is that the hard subprocess $M_H$ in eq.~(\ref{factor}) is approximated by
 the product of three quark amplitudes $M^m$.
 For $q q$ scattering both $t$- and  $u$- channels
 are available, fig.~2~(a,b),  and for $\qbar q$  there are
 $t$- and $s$- channels, fig.~2~(c,d).
 Given the above classification, there are four channel combinations
 that contribute to $p p$ or $n p$ elastic scattering,
 namely the direct $(ttt)$, fig.~3.(a), single interchange
 $(tuu +{\mathrm permutations}) $,
 fig.~3.(b), double interchange $(tuu + {\mathrm permutations})$
 and total interchange $(uuu)$.
 Similarly for $\pbar p$ the four possible combinations are
 obtained from the above by crossing from interchange to annihilation
 channels \mbox{($u \rightarrow s$)}.

 The color structure of the quark scatterings can be decomposed along
 a two dimensional color flow basis $\left( c_I \right)_{ \{a_i \} }$,
 $I=1,2$. For $q q \rightarrow q q$ we choose the basis
\beq
\left( c_1 \right)_{ \{a_i \}}  = \delta _{a_1 a_4} \delta_{a_2 a_3} \, ,
\hspace{1cm}
\left( c_2 \right)_{ \{a_i \}} = \delta _{a_1 a_3} \delta_{a_2 a_4} \, .
\label{ppbasis}
\eeq
At lowest order in $\alpha_s$ the color decomposition of the
direct and interchange channels is
\beqa
& & \left( C_t \right)_{ \{ a_i \} }
= (T_m)_{a_3 a_1} (T_m)_{a_4 a_2}
= A_1 \left(  c_1 \right)_{ \{a_i \}}
 +A_2 \left( c_2 \right)_{ \{a_i \}} \, ,
\nonumber \\
& & \left( C_u \right)_{ \{ a_i \} }
= (T_m)_{a_4 a_1} (T_m)_{a_3 a_2}
= A_2\left( c_1 \right)_{ \{a_i \} }
 +A_1 \left( c_2 \right)_{ \{a_i \}}   \, .
\label{tucolor}
\eeqa
 The color matrices $T_m$ are normalized as
 ${\mathrm tr} (T_m T_n) = (1/2) \delta_{m n}$ and the color decomposition
 coefficients are
\beq
A_1 = - N_c A_2 = \frac{1}{2} \, .
\label{Ccoeff}
\eeq
 For $\qbar q \rightarrow \qbar q$ we choose the $u \leftrightarrow s$
 crossed basis
\beq
\left( \cbar_1 \right)_{ \{ a_i \} }= \delta _{a_1 a_2} \delta_{a_3 a_4} \, ,
\hspace{1cm}
\left( \cbar_2 \right)_{ \{ a_i \} }= \delta _{a_1 a_3} \delta_{a_2 a_4} \, ,
\label{pbarpbasis}
\eeq
and the lowest order color decomposition of the  direct and annihilation
channels is
\beqa
& &\left( \Cbar_t \right)_{ \{ a_i \} }
= (T_m)_{a_1 a_3} (T_m)_{a_4 a_2}
=  A_1 \left( \cbar_1  \right)_{ \{ a_i \} }
  +A_2 \left( \cbar_2  \right)_{ \{ a_i \} }\, ,
\nonumber \\
& &\left( \Cbar_s \right)_{ \{ a_i \} }
= (T_m)_{a_1 a_2} (T_m)_{a_3 a_4}
= A_2 \left( \cbar_1  \right)_{ \{ a_i \} }
 +A_1 \left( \cbar_2 \right)_{ \{ a_i \} } \, ,
\label{stcolor}
\eeqa
 with coefficients as in eq.~(\ref{Ccoeff}).

 Since the amplitudes will be given in the helicity basis
 we state here the two approximations made concerning helicity.
 The first is that the total baryon helicity is the sum
 of the helicities of the valence constituents,
 eqs.~(\ref{helspinors}, \ref{tildeYexpand}).
 This is a consequence of leading twist factorization. Transverse
 momenta of the valence quarks are neglected relative to the
 hard scale ${\mathcal O}(|t|)$ and the quarks are assumed
 to be almost collinear and moving in the direction of the parent
 hadron.
 The second approximation is helicity conservation in the quark
 amplitudes $M^m, \: m=1,2,3$, up to ${\mathcal O}(m_q/ \sqrt{|t|})$
 corrections which can be neglected for light constituents and high
 momentum transfers. Then the quark amplitudes are scaleless and
 depend only on the CM angle $\theta$.
 Due to helicity conservation at the baryon and quark level
 the only non-vanishing helicity baryon amplitudes are
 \beqa
 & &\A(++;++) = \A(--;--) \, , \nonumber \\
 & &\A(+-;+-) = \A(-+;-+) \, ,  \\
 & &\A(+-;-+) = \A(-+;+-) \, . \nonumber
 \label{Ahel}
 \eeqa
 The $q q$ helicity Born amplitudes \cite{GastWu} for the $t$-channel are
 \beqa
 M_t(++;++) &=& -2 g^2 \frac{s}{t} C_t = 2 g^2
 \frac{1}{\sin^2(\theta/2)} C_t \, , \nonumber \\
 M_t(+-;+-) &=& 2 g^2 \frac{u}{t} C_t = 2 g^2
 \frac{\cos^2(\theta/2)}{\sin^2(\theta/2)} C_t \, , \\
 M_t(+-;-+) &=& 0 \, ,
 \nonumber
 \label{tborn}
 \eeqa
 and for the $u$-channel
 \beqa
 M_u(++;++) &=& - 2 g^2 \frac{s}{u} C_u =
 2 g^2 \frac{1}{\cos^2(\theta/2)} C_u \, , \nonumber \\
 M_u(+-;+-) &=& 0 \, , \\
 M_u(+-;-+) &=& 2 g^2 \frac{t}{u} C_u = 2 g^2
 \frac{\sin^2(\theta/2)}{\cos^2(\theta/2)} C_u \, .
 \nonumber
 \label{uborn}
 \eeqa
 The $\qbar  q$ $t$-channel amplitudes are as in eqs.~(19)
 but with opposite sign and the annihilation channel
 is the \mbox{$s \leftrightarrow u$} crossed version of
 eqs.~(20), i.e.
 \beqa
 M_s(++;++) &=& 0 \, , \nonumber \\
 M_s(+-;+-) &=& - 2 g^2 \frac{u}{s} \Cbar_s =
 2 g^2 \cos^2(\theta/2) \Cbar_s \, , \\
 M_s(+-;-+) &=&  2 g^2 \frac{t}{s} \Cbar_s =
 -2 g^2 \sin^2(\theta/2) \Cbar_s \, . \nonumber
 \label{sborn}
 \eeqa

 So far we have presented all the structures that determine
 the hadronic elastic amplitude at lowest order. Before giving
 its factorized form we discuss the effect of radiative corrections.
 In the formalism of Botts and Sterman \cite{BS}
 these are divided into two sets.
 The first set contains gluon exchange among quarks in the same
 hadron, which is factored into the hadronic wave functions.
 The second set contains soft gluon exchange among quarks from
 different hadrons, i.e. wave function irreducible corrections.
 These are factored into the color mixing tensor
 $U_{ \{a_i b_i c_i
      a^\prime_i b^\prime_i c^\prime_i \} }(\b_1,\b_2)$,
 which has the perturbative expansion
\beq
 U = \prod_{i=1}^{4}
\delta_{a_i a^\prime_i}\delta_{b_i b^\prime_i}\delta_{c_i c^\prime_i}
 + {\mathcal O} (\alpha_{s}(1/\b_m)) \, .
\label{U}
\eeq
 Primed indices are the color indices of the quark lines entering the
 hard scatterings and unprimed are the color indices of the quarks at
 hadronization, fig.~3.
 Therefore, the quark amplitudes carry primed color indices and the
 hadronic wavefunctions unprimed ones.
 At lowest order ${\mathcal O}(\alpha_s^0)$, $U$,
 denoted $U^{(0)}$ below,
 simply describes the absence of soft gluon exchange.

 Radiative corrections lead to logarithmic dependence on
 $s/ \mu^2$, $t/ \mu^2$ and $\b_m^2 / \mu^2$,
 where $\mu$ is the factorization scale. Logarithmic corrections
 can be resummed into exponential factors $\exp(-S_I)$, the Sudakov
 suppression factors.
 The Sudakov exponent $S_I$ corresponding to a certain
 hard scattering $M^m$ with color flow along the direction $I=1,2$,
 is \cite{Mueller,BS}
 \beq
 S_I(Q_m, \b_m) = \frac{8 C_F}{9} \ln (Q_m/ \Lambda)
 \ln \frac{ \ln ( Q_m / \Lambda)}{\ln (1/ |\b_m| \Lambda) }
 + ({\mathrm NL})_I \, ,
 \label{s}
 \eeq
 where $Q_m^2 = {\mathcal O}(x_m^2 |t|)$ is the hard scale of $M^m$ and
 $\Lambda$ is the QCD  scale parameter.
 In the axial gauge
 the leading logarithmic corrections describe the perturbative
 evolution of the wave function with the hard scale $Q$
 and the non-leading logarithmic corrections $({\mathrm NL})_I$
 are generated by the wave function irreducible soft gluon exchange
 and depend on the color flow $I$ of the hard scattering.

 We introduce the following notation.
 Given color tensor $C_{\{a_i, b_i c_i \}}$, $i=1,2,3,4$ we denote by
 ${\mathrm tr }_c (C)$ the contraction of the indices of $C$
 with the color structure of the baryon wave functions, i.e.
 \beq
 {\mathrm tr}_c (C) \equiv  \prod _{i=1}^4 \sum_{a_i,b_i c_i}
 \epsilon_{a_i b_i c_i} C_{ \{a_i, b_i, c_i \} } \, .
 \label{trdef}
 \eeq
 Then the factorized form of the elastic amplitude in
 impact parameter space is \cite{BS}
 \beq
 \A(s,t;h_{i}) =  \frac{N}{stu} \sum_f
  \int_{0}^{1} \frac{\d x_1 \d x_2}{ x_1^2 x_2^2 x_3^2}
 \int \d \b_1 \d \b_2  \R^{(f)} {\mathrm tr}_c (U M^1 M^2 M^3)
 \exp\left ( -S^1 -S^2 -S^3  \right )  \, ,
 \label{A}
 \eeq
 where
 \beq
 N= \frac{ \pi^6}{2 (N_c!)^4}
 \label{N}
 \eeq
 is a numerical constant depending on the normalization
 convention for the wave function, eq.~(\ref{tildeYexpand}),
 and $\sum_f$ takes into account all the quark scattering channel
 combinations.
 The channel index in the quark amplitudes $M^m$ has been left implicit.
 $\R^{(f)}$ is a fourth degree homogeneous polynomial
 of the hadronic wave functions $\P$.

 \section{ Color mixing and randomization}

 The decomposition of the color mixing tensor $U$ in the basis
 $\left( c_I \right)_{ \{a^\prime_i \} }$ defined in eq.~(\ref{ppbasis}) is
 \beq
 U_{IJK} \equiv {\mathrm tr}_c (U c_I c_J c_K) \, .
 \label{Udef}
 \eeq
 Since the color structure of the quark amplitudes factorizes from
 their helicity dependence we can readily
 separate the color coefficients through which each channel
 combination contributes to the amplitude $\A$
 in eq.~(\ref{A}) and express them in terms of the above color mixing tensor.
 For the direct channel the color coefficient is
 \beqa
 B_{ttt} &=& {\mathrm tr}_c(U C_t C_t C_t)  \nonumber \\
         &=& A_1^3 U_{111} + A_2^3 U_{222} + 3 A_1^2 A_2 U_{112}
             + 3 A_1 A_2^2 U_{122} \, ,
 \label{direct}
 \eeqa
 for the single interchange it is
 \beqa
 B_{ttu} &=& {\mathrm tr}_c(U C_t C_t C_u)  \nonumber \\
         &=&  A_1^2 A_2 U_{111} + A_1 A_2^2 U_{222}
     + \left( A_1^3 + 2 A_1 A_2^2 \right) U_{112}
     + \left( A_2^3 + 2 A_1^2 A_2 \right) U_{122} \, ,
 \label{single}
 \eeqa
 and for double and total interchange they are
 \beq
 B_{tuu} = B_{ttu}|_{A_1 \leftrightarrow A_2} \, ,
 \hspace{1cm}
 B_{uuu} = B_{ttt}|_{A_1 \leftrightarrow A_2} \, .
 \label{double}
 \eeq
 The above expressions distinguish explicitly
 between the color structure of the hard scattering contained in
 $A_1,\: A_2$ and the color mixing factors $U_{IJK}$ generated by
 soft gluon exchange.
 This distinction is important for the model we present below.

 The lowest order color mixing tensor $U^{(0)}$ has decomposition
 \beq
  U^{(0)}_{111} =U^{(0)}_{222}=36 \, ,
 \hspace{1cm}
  U^{(0)}_{112} = U^{(0)}_{221} = \frac{U^{(0)}_{222}}{3} \, ,
 \label{U0}
 \eeq
 which yields
 \beq
 B^{(0)}_{ttt} = B^{(0)}_{uuu} =\frac{10}{3} \, ,
 \hspace{1 cm}
 B^{(0)}_{ttu} =B^{(0)}_{tuu} = -\frac{2}{9} \, .
 \label{B0}
 \eeq
 Note that the single and double interchange coefficients are
 (-1/15) times the direct or total interchange.
 Higher order in soft gluon exchange coefficients $B^{(n)}$ can be
 constructed from the perturbative expansion of $U$
 to ${\mathcal O}(\alpha_s^n(1/\b_m))$.
 This expansion has a clear meaning in the asymptotic region
 \mbox{$ s \sim |t| \rightarrow \infty$}
  where the Sudakov suppression $\exp(-S)$
 forces the hard scatterings close together,
 so that $1/ \b_m$ can be treated  as a perturbative scale \cite{BS}.
 Botts \cite{Botts} however finds that
 the onset of asymptopia where the process
 is dominated by these perturbative contributions occurs at very high
 energies, $\ln(s/s_0) \sim 8 \,, \: s_0 = 1{\mathrm GeV}^2$.
 This suggests that at moderate energies, where measurements are
 available, the Sudakov suppression becomes far less effective
 and $U$ is beyond perturbative control.
 We suggest that the effect of soft gluon exchange in this region  is to
 decorrelate the color configurations of the constituent quarks in
 the initial and final state from the color configurations they
 have when they participate in the hard scatterings. In other
 words, due to strong color mixing,
 the color indices of the quark lines, fig.~3,
 have been randomized by $U$ by the time they enter the hard scatterings.
 We build this into the formalism  by requiring the color mixing tensor
 $U_{IJK}$ to be totally symmetric in the
 bases of eqs.~(\ref{ppbasis},\ref{pbarpbasis}) and to satisfy
 \beq
 U^{\mathrm rand}_{111}=U^{\mathrm rand}_{222}
 =U^{\mathrm rand}_{112}=U^{\mathrm rand}_{221} \, .
 \label{decorrelation}
 \eeq
 Compare this with eq.~(\ref{U0}) and
 note that the above relation is not assumed
 to be valid order by order in $\alpha_s$.
 It is a statement about color flow in the non-perturbative regime.
 The color randomization condition (\ref{decorrelation}) yields
 via eqs.~(\ref{direct}-\ref{double})
 \beq
 B^{\mathrm rand}_{ttt} = B^{\mathrm rand}_{uuu} =
 B^{\mathrm rand}_{ttu} = B^{\mathrm rand}_{tuu} \, .
 \label{wierd}
 \eeq
 This relation holds independently of the specific value of the
 hard color coefficients $A_1, A_2$. It means that
 the color structure of the short distance subprocess becomes
 irrelevant for the determination of the hadronic amplitude
 exactly because it is unstable under soft gluon exchange
 over large space-time
 scales.\footnote[1]{
 Similar ideas have been suggested in the context of diffractive
 D.I.S. and heavy onium production in refs.
 \cite{Buchmuller} and \cite{Halzen} respectively. }
 Since all channel configurations contribute with the same color coefficients
 in the Landshoff mechanism, the elastic scattering will be dominated
 by the interchange channels as they are more numerous.

 In order to demonstrate the combined effect of flavor flows
 and color factors without  the complications of spin we
 consider a toy model where the constituent quarks are scalars.
 The quark Born amplitudes, fig.\ 2, are
 \beqa
 & & M_t = g^2  \frac{1}{t} (k_1+k_3) \cdot (k_2+k_4)
 \left(C_t\right)_{ \{ a_i \}}
 = g^2 \left( \frac{s-u}{t} \right) \left( C_t \right)_{ \{ a_i \}} \, ,
 \nonumber \\
 & & M_u = g^2  \frac{1}{u} (k_1+k_4) \cdot (k_2+k_3)
 \left(C_u\right)_{ \{ a_i \}}
 = g^2 \left( \frac{s-t}{u} \right) \left( C_u \right)_{ \{ a_i \}}\, ,
 \nonumber \\
 & & M_s = g^2  \frac{1}{s} (k_1-k_2) \cdot (k_3-k_4)
 \left( \overline{C}_s\right)_{ \{ a_i \}}
 = g^2 \left( \frac{u-t}{s} \right) \left( \Cbar_s \right)_{ \{ a_i \}} \, .
 \label{scalarBorn}
 \eeqa
 The $p p$ and $\pbar  p$  amplitudes are determined by simply
 counting the available channel combinations.
 \beq
 \A^{pp}_{\mathrm scalar} =
 \frac{N}{s t u} {\mathrm tr}_c
 \left[ U [ 3 (M_t+M_u)^3 + 6 (M_t^2 + M_u^2) (M_t + M_u) ] \right]
 {\mathcal I} \, ,
 \label{ppscalar}
 \eeq
 \beq
 \A^{\pbar p}_{\mathrm scalar} =
 \frac{N}{s t u} {\mathrm tr}_c
 \left[ U [ 3 (M_t+M_s)^3 + 6 (M_t^2 + M_s^2)( M_t+M_s) ] \right]
 {\mathcal I} \, .
 \label{pbarpscalar}
 \eeq
 The first term in eq.~(\ref{ppscalar}) comes from the contributions
 of fig.~1~(a) and the second from fig.~1~(b,c).
 The flavor inequivalent reorderings of the hard scatterings
 have been taken into account.
 The factor ${\mathcal I}$  contains the integrations over momentum fractions
 and impact parameters as in eq.~(\ref{A}) and depends on the
 hadronic mass scales.
 At $\theta = 90^0$, annihilation channels do not contribute
 ($M_s = 0$) and direct and interchange channels are equal up to
 their respective color structure.
 Then the amplitude ratio depends only on the color
 coefficients $B$
 \beq
 \left| \frac{\A^{\pbar p}}{\A^{pp}}
 \right| _{\theta=90^0} =
 \left| \frac{ 9 B_{ttt}} {9 B_{ttt} + 9B_{uuu} + 15 B_{ttu}
        + 15 B_{tuu}} \right| \, .
 \label{Ampratio}
 \eeq
 At asymptotically high energies the lowest perturbative order
 coefficients $B^{(0)}$ of eqs.~(\ref{B0}) give
 \beq
 R^{\mathrm as}_{\pbar p}(s, \theta=90^0) =
 \left( \frac{9}{16} \right)^2 =\frac{1}{3.16} \, .
 \label{scalarratioasy}
 \eeq
 At moderate energies, where color randomization is assumed to occur,
 the color coefficients $B^{\mathrm rand}$, eq.~(\ref{wierd}), yield
 \beq
 R^{\mathrm rand}_{\pbar p}(s, \theta=90^0) =
 \frac{1}{9} \, R^{\mathrm as}_{\pbar p}(s, \theta=90^0) =
 \frac{1}{28.4} \, .
 \label{scalarratioran}
 \eeq
 Color randomization results in a smaller ratio
 because the interchange channels in eq.~(\ref{ppscalar})
 give much bigger relative contribution to the $p p$ amplitude
 than in lowest order in pQCD.
 In the next section we shall see that this feature of Landshoff
 scattering persists after the inclusion of the spin of the quarks.

 \section{ Elastic scattering in the helicity basis }

 In this section we compute the baryon-baryon ($p p$, $\pbar p$ and
 $n p$) elastic amplitude for wide-angle scattering
 both in the asymptotic energy limit and for moderate energies, where
 the assumption of color randomization is believed to be relevant.
 According to eq.~(\ref{A}) the elastic amplitude for given baryon
 helicities is obtained by summing over all quark scattering channels
 that are allowed by helicity conservation, weighed by the appropriate
 wave function factors $\R$ and color traced after
 contraction with the color mixing tensor $U$.
 For very large momentum transfer $|t|$ the dependence of the
 Sudakov exponent $S_I$ on the color flow of the corresponding
 hard scattering enters as non-leading logarithmic dependence
 on $t$, eq.~(\ref{s}).
 It has been argued that these non-leading logarithmic corrections
 can give rise to non-trivial phase structure in the amplitude
 that may account for its oscillatory behavior with energy
 \cite{Pire, Carlson}.
 In the following we are going to neglect them because,
 although our model retains the flavor and crossing structure
 of pQCD, it is suggested to be valid in an energy region where
 the perturbative expansion of radiative corrections is not
 applicable. In this case we will actually set the whole Sudakov exponent
 equal to zero.

 The decomposition of the hadronic state in terms of quark helicities
 is given by eq.~(\ref{tildeYexpand}).
 The results are given in terms of a general ${\cal P}_{123}$,
 whose explicit form is determined by considering specific models
 for the  proton wave function  \cite{SS2,LCDA}.
 The $p p$ and $\pbar p$ amplitudes can be expressed in terms of the
 following five wave function combinations.
 \noindent
 \beqa
 & &\R_0 = \P^4_{123} + \P^4_{213}
 + 16 \T_{123}^4 + (1 \leftrightarrow 3) + (2 \leftrightarrow 3) \, ,
 \nonumber \\
 & &\R_1 = 2 \P^2_{123} \P^2_{213}
               + 8 \P^2_{123} \T^2_{123}
               + 8 \P^2_{213} \T^2_{123}
               + (1 \leftrightarrow 3) + (2 \leftrightarrow 3)\, ,
 \nonumber \\
 & &\R^\prime_1 =
 2 \P^2_{123} \P^2_{312} +
 2 \P^2_{132} \P^2_{213} +
 8 \P^2_{213} \T_{132}^2 +
 8 \P^2_{312} \T_{123}^2
  + (1 \leftrightarrow 2) + (1 \leftrightarrow 3) \, ,
 \nonumber \\
 & &\R^\prime_2 =
 2 \P^2_{123} \P^2_{132} +
 32 \T^2_{123} \T^2_{132}
  + (1 \leftrightarrow 2) + (1 \leftrightarrow 3) \, ,
 \\
 & &\R^\prime_3 =
 16 \P_{123} \P_{132} \T_{123} \T_{132}
  + (1 \leftrightarrow 2) + (1 \leftrightarrow 3) \, .
 \nonumber
 \label{R}
 \eeqa
 The additional terms generated by the permutations shown
 in the above equations are due to the flavor inequivalent
 relabelings of the three hard scatterings $M^m$.
 $\R_1$  contributes to the diagram fig.~1~(a),
 the three $\R^\prime$'s contribute to the diagrams fig.~1~(b,c),
 and $\R_0$ contributes to all three diagrams fig.~1~(a,b,c). \\
 The $p p$ helicity amplitudes are
 \beqa
 & &\A^{pp}(++;++) = -\frac{N (8 \pi)^3}{s t u}
 \int_0^1 \frac{\d x_1 \d x_2}{x_1^2 x_2^2 x_3^3}
 \int \d \b_1 \d \b_2 \alpha_s^3(\mu) \exp(-S^1 -S^2 -S^3)
 \nonumber \\
 & & \hspace{1cm} \times
  \left[ B_{ttt} (2 \R_0)
       \left( \frac{s^3}{t^3} + \frac{s^3}{u^3} \right)
       +B_{ttt} (\R_1 + \R_1^\prime + \R_2^\prime)
       \left( \frac{s u^2}{t^3} + \frac{s t^2}{u^3} \right) \right.
 \nonumber \\
 & & \hspace{1.5cm} \left.
  +B_{ttu} (4 \R_0)
       \left( \frac{s^3}{t^2 u} +\frac{s^3}{t u^2} \right)
 + B_{ttu} (\R_1 + \R_2^\prime)
    \left( \frac{s u}{t^2} + \frac{s t}{u^2} \right) \right]
 \, ,
 \label{pp1}
 \eeqa
 \beqa
 & &\A^{pp}(+-;+-) = \frac{N (8 \pi)^3}{s t u}
 \int_0^1 \frac{\d x_1 \d x_2}{x_1^2 x_2^2 x_3^3}
 \int \d \b_1 \d \b_2 \alpha_s^3(\mu) \exp(-S^1 -S^2 -S^3)
 \nonumber \\
 & & \hspace{1cm} \times
  \left[ B_{ttt} (2 \R_0) \frac{u^3}{t^3}
  + B_{ttt} (\R_1 + \R_1^\prime + \R_2^\prime) \frac{s^2 u}{t^3} \right.
 \nonumber \\
 & &  \hspace{1.5cm} \left.
 + B_{ttu}(2 \R_1 + \R_1^\prime) \frac{s^2}{t^2}
  + B_{ttu}(\R_1 + \R_3^\prime)
  \left( \frac{t}{u} +\frac{s^2}{t u} \right) \right]
 \, ,
 \label{pp2}
 \eeqa
 and
 \beq
 \A^{pp}(+-;-+) = \A^{pp}(+-;+-)
                                 |_{ t\leftrightarrow u} \, .
 \label{pp3}
 \eeq
 The $\pbar p$ helicity amplitudes are obtained from the
 above $p p$ amplitudes via the following crossings.
 \beqa
 \A^{\pbar p}(++;++) &=& \A^{pp}(+-;+-)
 | _{s \leftrightarrow u} \, ,
 \nonumber \\
 \A^{\pbar p}(+-;+-) &=& \A^{pp}(++;++)
 | _{s \leftrightarrow u} \, ,
 \\
 \A^{\pbar p}(+-;-+) &=& \A^{pp}(+-;-+)
 | _{s \leftrightarrow u} \, .
 \nonumber
 \label{pbarp}
 \eeqa
 Finally, the $n p$ helicity amplitudes are given in the appendix.

 The observables we are considering here do not depend significantly
 on the specific form of the hadronic wavefunction. Model ligh-cone
 distribution amplitudes affect the overall normalization of the
 hadronic amplitudes because of the asymmetric distribution
 of the longitudinal momentum among the valence quarks
 \cite{CZrep, LCDA}.
 Due to the permutations of the arguments in eqs.~(41), though,
 the wave function model dependence of the cross section ratios and the
 angular distribution is expected to be minimal \cite{Botts}.
 Consequently,
 to compute $R_{\pbar p/ pp} $ we use the totally symmetric
 \mbox{${\cal P}_{123} = {\cal P}_{\rm as}$},
 and the above wave function combinations become
 \beqa
 & &\R_0 = 54 {\cal P}_{\rm as}^4 \, ,
 \hspace{.5cm}
 \R_1 = 54 {\cal P}_{\rm as}^4 \, ,
 \nonumber \\
 & &\R^\prime_1 = 60  {\cal P}_{\rm as}^4 \, ,
 \hspace{.5cm}
 \R^\prime_2 = 102  {\cal P}_{\rm as}^4 \, ,
 \hspace{.5cm}
 \R^\prime_3 = 48  {\cal P}_{\rm as}^4 \, .
 \label{Ras}
 \eeqa
 The above form for the wave function combinations
 and eqs.~(42-45) reproduce the results of
 Farrar and Wu in ref.~\cite{FarrarWu} up to
 an overall normalization factor.
 The ratio $R_{\pbar p/pp}$ is given in the helicity basis by
 \beq
  R_{\pbar p/pp} = \frac{ |\A^{\pbar p}(++;++)|^2
+ |\A^{\pbar p}(+-;+-)|^2+ |\A^{\pbar p}(+-;-+)|^2}
  {|\A^{p p}(++;++)|^2
 +|\A^{p p}(+-;+-)|^2+|\A^{p p}(+-;-+)|^2} \, .
 \label{Rform}
 \eeq
 Using the lowest perturbative order color coefficients $B^{(0)}$
 in eq.~(\ref{B0}) we obtain the result
 \beq
 R^{\mathrm as}_{\pbar p/pp}(s, \, \theta=90^0) \approx \frac{1}{2.68} \, .
 \label{ppratioas}
 \eeq
 This result, although definitely less than unity, is
 much larger than the experimental value
 \mbox{$R_{\pbar p/pp} \approx 1/40$},
 measured at \mbox{$\sqrt{s} = 3.59$ GeV} \cite{AGS}.
 For the color randomization model, we compute $R_{\pbar p/pp}$ using again
 the asymptotic wave functions of eq.~(\ref{Ras}),
 and the color factors $B^{\mathrm rand}$ of eq.~(\ref{wierd}).
 The result is
 \beq
 R^{\mathrm rand}_{\pbar p/pp}(s, \, \theta=90^0) \approx \frac{1}{27.7} \, ,
 \label{ppratio}
 \eeq
 Color randomization yields a ratio one order of magnitude smaller
 than the asymptotic case and close in value to the scalar quark toy
 model of the previous section.
 The corresponding results for $np$ elastic scattering are obtained
 by using the helicity amplitudes given in the appendix.
 \beq
 R^{\mathrm as}_{np/pp}(s, \, \theta=90^0) \approx 0.30 \, ,
 \hspace{1cm}
 R^{\mathrm rand}_{np/pp}(s, \, \theta=90^0) \approx 0.36 \, .
 \label{npratioas}
 \eeq
 Color randomization gives slightly bigger ratio for $np/pp$
 elastic scattering unlike the case of $\pbar p/pp$.
 Both values of $R_{np /pp}$
 though are compatible with the experimental value  \cite{Stone}
 $R_{np/pp} = 0.34 \pm 0.05$ measured over an energy range \\
 3.10~GeV~$\le~\sqrt{s}~\le$~4.75~GeV.

 Finally we examine the effect of color randomization on the angular
 distribution of $p p$ elastic scattering.
 To this end we plot the differential cross section normalized
 at $\theta=90^0$ versus $\cos \theta$, fig. 4. Landshoff scattering
 in the asymptotic limit yields a steeply rising angular distribution,
 approximately of the form $(1-\cos^2\theta)^{-12}$.
 Color randomization softens this distribution to an approximate
 form $(1-\cos^2\theta)^{-10}$. This is to be compared with
 the fit to the experimental data $(1-\cos^2\theta)^{-7}$ given
 by Farrar and Wu in \cite{FarrarWu}. In all cases the angular
 distribution is independent of the CM energy.
 The color randomization distribution is in relative good agreement
 with the experimental fit for \mbox{$\cos(\theta) \le 0.3$} but it
 becomes much steeper away from the central region.

 \section{Summary}

 We have considered wide-angle elastic scattering in the Landshoff
 mechanism and organized the calculation making explicit the
 effect of color.
 For scattering at moderate energies
 we have suggested a pQCD motivated model which realizes the
 observation that the elastic scattering is dominated by
 quark interchange among the hadrons.
 This is assumed to occur because the color of the constituent quarks
 is totally randomized by soft gluon exchange.
 By implementing this in the expressions for the hadronic helicity
 amplitudes we obtain a cross section ratio
 $R_{\pbar p/ pp}$ which is an order of magnitude smaller than
 the asymptotic value and compatible with the experimental measurements
 \cite{Blazey, AGS}. This feature of  $R_{\pbar p/ pp}$ is due to the
 nature of the Landshoff mechanism. Because in this picture the elastic
 process occurs via independent quark scattering, a change in the
 relative contribution between channels at the quark level has a
 sizable effect in the hadronic cross section.
 Color randomization leads to softening of the angular distribution
 in $p p$ scattering, although we found that
 away from the central region
 it is still steeper than what experiment suggests.

 On the theoretical side, the separation of
 gluons into hard and soft becomes less clear away from asymptopia,
 due to the small momentum transfers involved.
 Moreover, conservation of color and color randomization require
 to include components of the hadronic wave function
 beyond the leading twist three quark part.
 Another set of approximations we made
 has to do helicity conservation.
 In the moderate energy regime, quark mass and intrinsic transverse
 momentum  corrections can be important.
 This is the reason why we did not reproduce
 the $\sim s^{-10}$ scaling of the elastic cross section.
 We considered instead observables which are less sensitive
 to the specific form of the hadronic wavefunction
 or the factorization assumptions and
 mainly determined by the flavor routing of the constituent quarks.
 It would be of interest to analyze the contribution of the
 Landshoff mechanism relative to the QIM mechanism,
 as in ref \cite{RamseySivers2},
 but taking  into account color randomization.

\hspace{2cm}

 {\em Acknowledgements \/} The author would like to thank George Sterman
 for many insightful discussions and suggestions.

 \newpage

 \appendix

 \section{Appendix}

 \setcounter{equation}{0}
 \renewcommand{\theequation}{\Alph{section}.\arabic{equation}}

 Due to isospin symmetry the wave function for the neutron is
 obtained from the corresponding one for the proton via the
 substitution $ u \rightarrow -d$ and $d \rightarrow u$ in
 eqs.~(\ref{helspinors},~\ref{tildeYexpand}).
 For $n p$ elastic scattering, apart from the wave function
 combinations given in eq.~(\ref{R}), an additional one is needed,
 namely
 \beq
 \R^\prime_4 = 8 \P_{123} \P_{213} \P_{312} \T_{132} +
               8 \P_{132} \P_{312} \P_{213} \T_{123}
           + (1 \rightarrow 2) + (1 \rightarrow 3) \, .
 \label{R4}
 \eeq
 The $n p$ helicity amplitudes are
 \beqa
 & &\A^{np}(++;++) = -\frac{N (8 \pi)^3}{s t u}
 \int_0^1 \frac{ \d x_1 \d x_2}{x_1^2 x_2^2 x_3^3}
 \int \d \b_1 \d \b_2 \alpha_s^3(\mu) \exp(-S^1 -S^2 -S^3)
 \nonumber \\
 & & \hspace{1cm} \times
 \left[ B_{ttt} (2 \R_0)
  \left( \frac{s^3}{t^3} + \frac{s^3}{u t^2} + \frac{s^3}{t u^2} \right)
       +B_{ttt} (\R_1+\R_1^\prime+\R_2^\prime) \frac{s u^2}{t^3} \right.
 \nonumber \\
 & &  \hspace{1.5cm} \left.
    +B_{ttu} \R_1^\prime \frac{s u}{t^2}
    + B_{ttu} (2 \R_3^\prime) \frac{s t}{u^2} \right]
 \, ,
 \label{np1}
 \eeqa
 \beqa
 & &\A^{np}(+-;+-) = \frac{N (8 \pi)^3}{s t u}
 \int_0^1 \frac{ \d x_1 \d x_2}{x_1^2 x_2^2 x_3^3}
 \int \d \b_1 \d \b_2 \alpha_s^3(\mu) \exp(-S^1 -S^2 -S^3)
 \nonumber \\
 & & \hspace{1cm} \times
 \left[ B_{ttt} (2 \R_0) \frac{u^3}{t^3}
       +B_{ttt} (\R_1 + \R_1^\prime + \R_2^\prime) \frac{s^2 u}{t^3} \right.
 \\
 & & \hspace{1.5cm} \left.
       +B_{ttu} (\R_1^\prime + 2 \R_2^\prime) \frac{s^2}{t^2}
       +B_{ttu} (\R_2^\prime + \R_3^\prime)
               \left( \frac{s^2}{t u} + \frac{t}{u} \right) \right] \, ,
 \label{np2}
 \eeqa
 and
 \beqa
 & &\A^{np}(+-;-+) = -\frac{N (8 \pi)^3}{s t u}
 \int_0^1 \frac{ \d x_1 \d x_2}{x_1^2 x_2^2 x_3^3}
 \int \d \b_1 \d \b_2 \alpha_s^3(\mu) \exp(-S^1 -S^2 -S^3)
 \nonumber \\
 & & \hspace{3cm} \times
  B_{ttu} \R^\prime_4 \left( 2 \frac{s^2}{u^2} + \frac{s^2}{t u} +
                             \frac{u}{t} \right)
 \, .
 \label{np3}
 \eeqa

 \newpage

 \begin{center}
 {\bf Figure Captions}
 \end{center}

 \begin{enumerate}
 \item  Proton-proton elastic scattering in the Landshoff mechanism
       and inequivalent flavor routings. The dashed lines
       represent the $d$-quarks. All momenta flow from left to right.

 \item Exchange channels for quark-quark (a,b) and antiquark-quark
       scattering (c,d).

 \item Soft gluon exchange and color mixing for the direct ($ttt$), (a),
       and the single interchange ($utt$) channel, (b),
       in baryon-baryon elastic scattering.
       Hard gluons are not shown.
       Interpreted as color graphs, these diagrams represent contributions
       to $U_{222}$, (a), and $U_{211}$, (b).

 \item Angular distribution for proton-proton elastic scattering.
       The data fit is from ref. \cite{FarrarWu}

 \end{enumerate}

 \newpage
 \begin{figure}
 \centerline{\psfig{figure=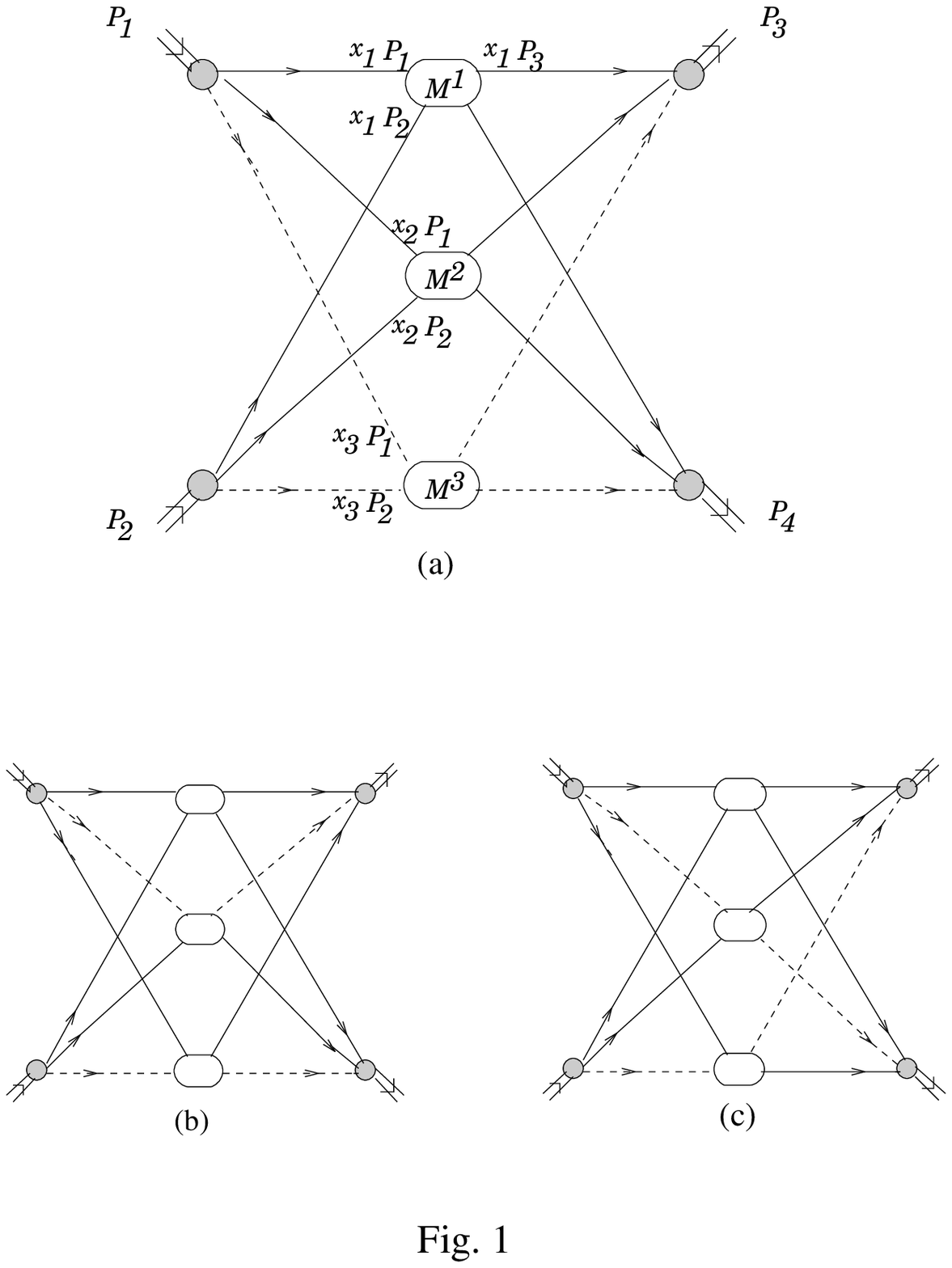}}
 \end{figure}

 \newpage
 \begin{figure}
 \centerline{\psfig{figure=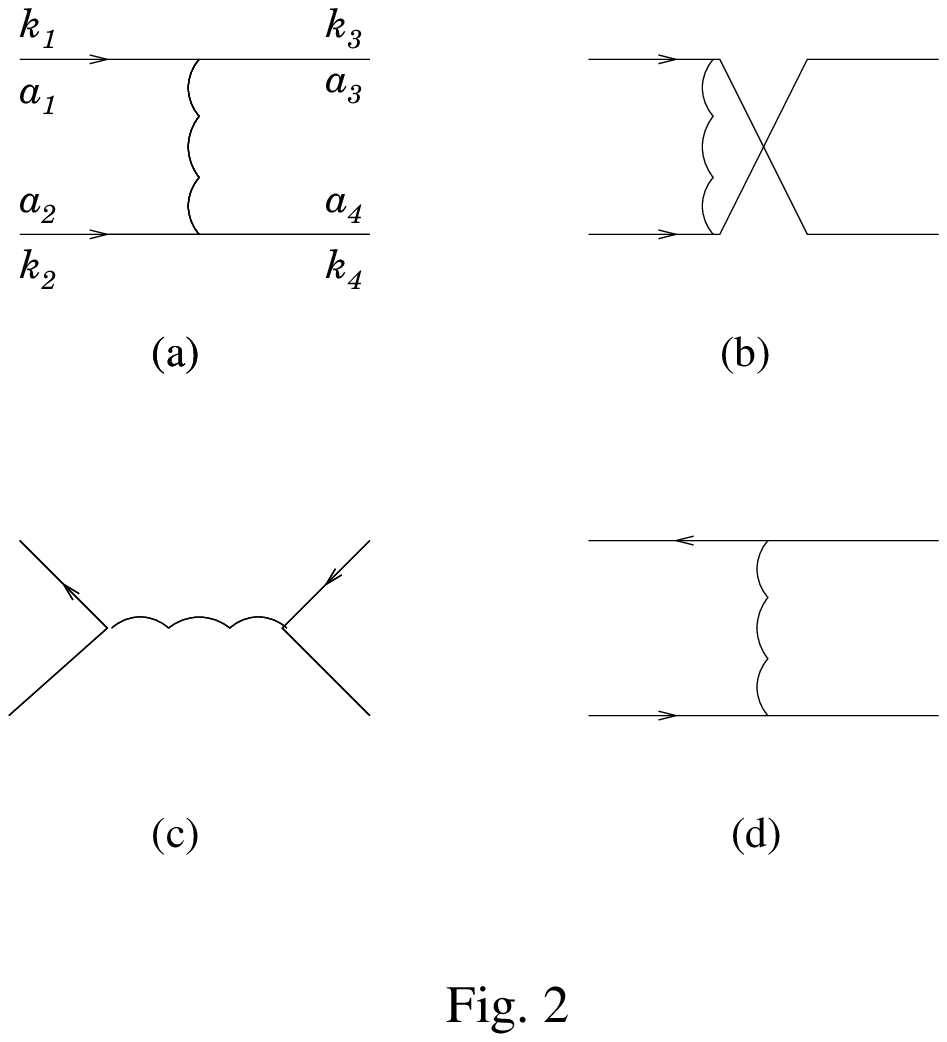}}
 \end{figure}

 \newpage
 \begin{figure}
 \centerline{\psfig{figure=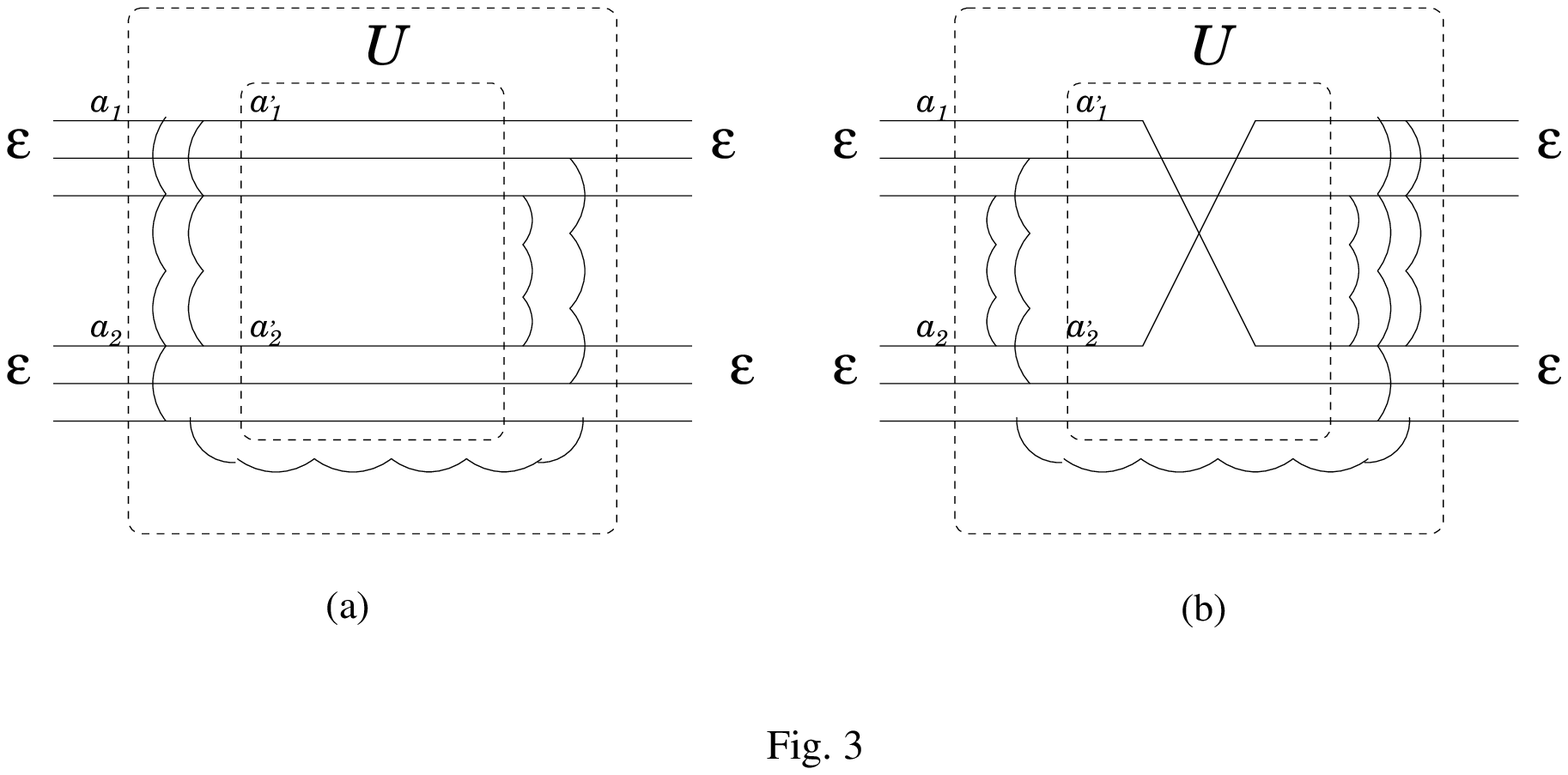}}
 \end{figure}

 \newpage
 \begin{figure}
 \centerline{\psfig{figure=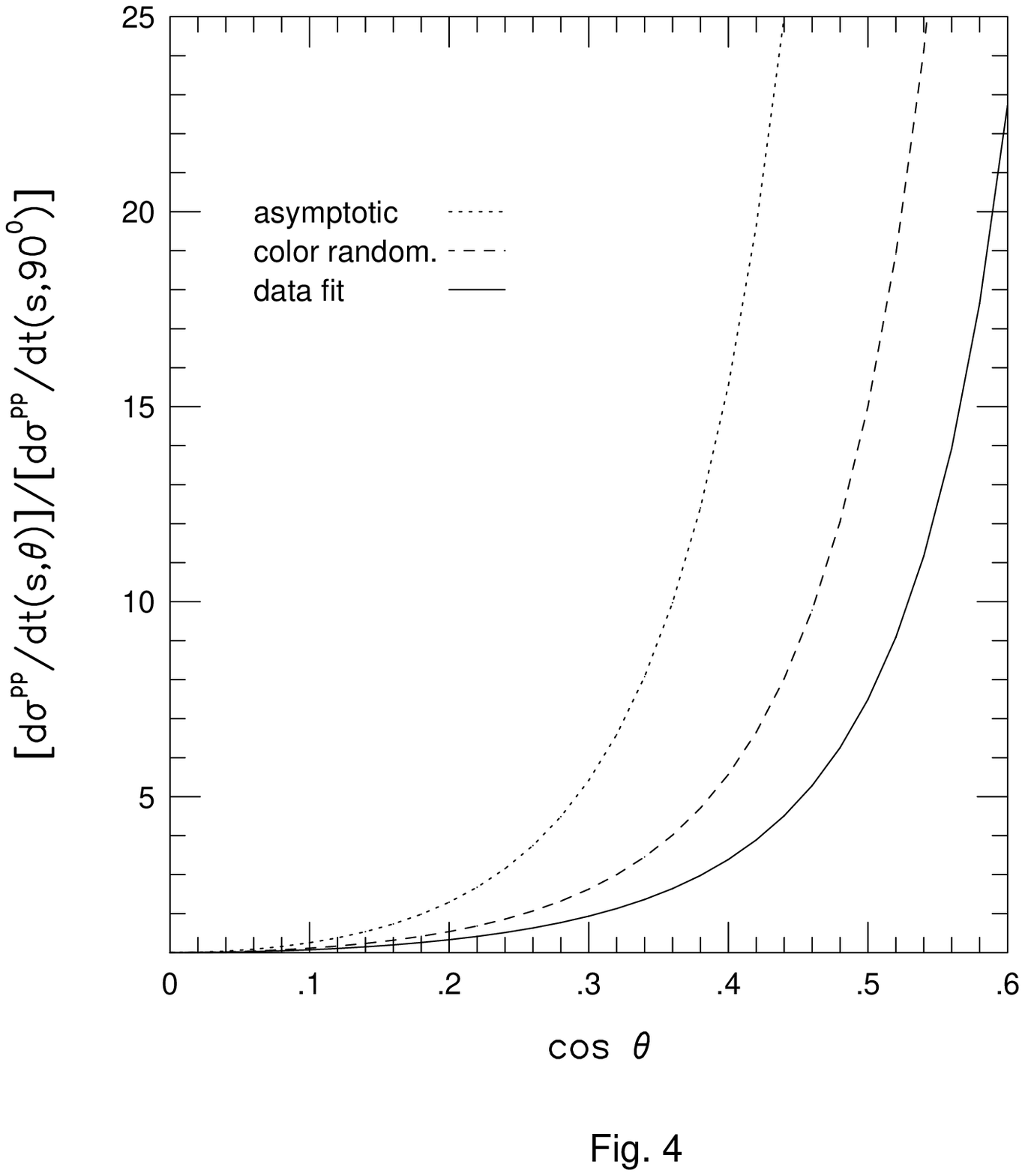}}
 \end{figure}


\newpage

 \begin{thebibliography} {99}


 \bibitem{BL}
   S.J. Brodsky and G.P. Lepage, in {\em Perturbative Quantum
     Chromodynamics},
  edited by  A. H. Mueller (World Scientific, Singapore, 1989)

 \bibitem{FarrarNeri}
   G. R. Farrar and F. Neri, Phys. Lett. \textbf{130B}, 109
          (1983); \textbf{152B} 443 (1985)

 \bibitem{Gunion} J. F. Gunion, S. J. Brodsky and R. Blankenbecler,
                Phys. Rev. D {\textbf 8}, 287 (1973); \\
                D. Sivers, S. J. Brodsky and R. Blankenbecler,
                Phys. Rep. C \textbf{23}, 1 (1976)

 \bibitem{Land} P. V. Landshoff, Phys. Rev. D \textbf{10}, 1024 (1974); \\
         A. Donnachie and P. V. Landshoff, Z. Phys. C \textbf{2}, 55 (1979)

 \bibitem{RamseySivers1} G. P. Ramsey and D. Sivers,
         Phys. Rev. D \textbf{45}, 79 (1992); D \textbf{47}, 93 (1993)

 \bibitem{Blazey} G. C. Blazey {\em et al.},
                 Phys. Rev. Lett. \textbf{55}, 1820 (1985);\\
  B. R. Baller {\em et al.}, Phys. Rev. Lett. \textbf{60}, 1118 (1988);\\
    A. Carrol {\em et al.}, Phys. Rev. Lett. \textbf{61}, 1698 (1988)

 \bibitem{AGS} C. White {\em et al.}, Phys. Rev. D \textbf{49}, 58 (1994)

 \bibitem{Botts} J. Botts, Nucl. Phys. \textbf{B353}, 20 (1991)

 \bibitem{CZrep} V. L. Chernyak and A. R. Zhitnitsky,
               Phys. Rep. \textbf{112}, 173 (1984)

 \bibitem{SS2} M. G. Sotiropoulos and G. Sterman,
              Nucl. Phys. \textbf{B425}, 489 (1994)

 \bibitem{LCDA} V. L. Chernyak and I. R. Zhitnitsky,
           Nucl. Phys. \textbf{B246}, 52  (1984); \\
          I. R. Zhitnitsky, A. A. Ogloblin and V. L. Chernyak,
         Yad. Fiz. \textbf{48}, 841 (1988)
         [Sov. J. Nucl. Phys. \textbf{48}, 536 (1988)]; \\
   I. D. King and C. T. Sachrajda, Nucl. Phys. \textbf{B279}, 785 (1987); \\
   M. Gari and N. G. Stefanis, Phys. Rev. D \textbf{35}, 1074 (1987)

 \bibitem{GastWu} R. Gastmans and T. T. Wu,  {\em The Ubiquitous
                 Photon} (Clarendon Press, Oxford, 1990)

 \bibitem{BS} J. Botts and G. Sterman, Nucl. Phys. \textbf{B325}, 62
             (1989)

 \bibitem{Mueller}   A. H. Mueller, Phys. Rep. \textbf{73}, 237 (1981)

 \bibitem{Buchmuller} W. Buchm\"{u}ller and A. Hebecker,
               DESY preprint, DESY 95-077, April 1995

 \bibitem{Halzen} J. F. Amundson, O. J. P. \'{E}boli, E. M. Gregores
   and F. Halzen, Madison preprint, MADPH-95-919, December 1995

 \bibitem{Pire} B. Pire and J. P. Ralston,
         Phys. Lett. \textbf{117B}, 233 (1982)

 \bibitem{Carlson} C. E. Carlson, M. Chachkhunashvili and F. Myhrer,
                  Phys. Rev. D \textbf{46}, 2891 (1992)

 \bibitem{FarrarWu} G. R. Farrar and C.-C. Wu,
                    Nucl. Phys. \textbf{B85}, 50 (1975)

 \bibitem{Stone} J. L. Stone {\em et al.},
       Phys. Rev. Lett. \textbf{23}, 1315 (1977); 1317 (1977)

\bibitem{RamseySivers2} G. P. Ramsey and D. Sivers,
                    Phys. Rev. D \textbf{52}, 116 (1995)

 \end{thebibliography}
 \end{document}